\documentclass[%
 reprint,
nofootinbib,
 amsmath,amssymb,
 aps,
]{revtex4-1}

\usepackage{graphicx}
\usepackage{dcolumn}
\usepackage{bm}
\usepackage{hyperref}

\usepackage{amsfonts}
\usepackage{setspace}
\usepackage{amssymb}
\usepackage{amsmath}
\usepackage{subfigure}
\usepackage{color}
\usepackage{longtable}

\def\Ri{{\cal R}}
\def\phit{\tilde{\phi}}

\def\hz{h_z}
\def\PL{P_\Lambda}
\def\Sa{S_1}
\def\Sb{S_2}
\def\Sc{R_{,y}}
\def\Sd{S_3}

\newcommand{\ba}{\begin{array}}
\newcommand{\ea}{\end{array}}
\newcommand{\be}{\begin{equation}}
\newcommand{\ee}{\end{equation}}
\newcommand{\bea}{\begin{eqnarray}}
\newcommand{\eea}{\end{eqnarray}}

\begin{document}


\title{
Modelling the Evaporation of Non-singular Black Holes}



\author{Tim Taves}
  \homepage{http://www.cecs.cl}
 \affiliation{%
Centro de Estudios Cient{\'i}ficos (CECs), Valdivia, Chile
 }%
\author{Gabor Kunstatter}
\homepage{http://ion.uwinnipeg.ca/~gkunstat}
\affiliation{%
  Physics Department, University of Winnipeg
}%


\date{\today}

\begin{abstract}

 We present a model for studying the formation and evaporation of non-singular (quantum corrected) black holes. The model is based on a generalized form of the dimensionally reduced, spherically symmetric Einstein--Hilbert action and includes a suitably generalized Polyakov action to provide a mechanism for radiation back-reaction.  The equations of motion describing self-gravitating scalar field collapse are derived in local form both in null co--ordinates and in Painleve--Gullstrand (flat slice) co--ordinates. They provide the starting point for numerical studies of complete spacetimes containing dynamical horizons that bound a compact trapped region. Such spacetimes have been proposed in the past as solutions to the information loss problem because they possess neither an event horizon nor a singularity.  Since the equations of motion in our model are derived from a diffeomorphism invariant action they preserve the constraint algebra and  the resulting energy momentum tensor is manifestly conserved.

\end{abstract}

\maketitle


\section{Introduction and Background}
General relativity predicts the  existence of singularities at the center of all black holes, while Hawking's famous quantum calculation implies that event horizons shrink by emitting thermal radiation. These two properties of black holes, combined with the fact that special relativity forbids information from escaping from an event horizon,  lead to one of the deepest puzzles of modern theoretical physics: the so-called information loss paradox. In its simplest form, the question is: What is the endpoint of black hole formation and evaporation and what happens to the information about the state of the matter that formed the black hole? 

A variety of end states have been suggested over the years. Among them:
\begin{enumerate}
\item The black hole evaporates completely via thermal radiation so that the information is lost to the outside world (asymptotic region of the black hole spacetime) forever. This scenario entails a breakdown of the unitarity of quantum mechanics.
\item At late stages the radiation is no longer thermal and the information emerges via quantum gravity related, presumably causality violating, corrections.
\item The black hole stops emitting radiation when it nears the Planck scale, leaving behind a microscopic remnant that hides forever an enormous amount of information.
\item Most recently it has been suggested \cite{Almheiri2013a,Almheiri2013}\footnote{See also \cite{Vaz2014,Vaz2014a} where this issue is addressed in a canonically quantized model of dust collapse.} that a firewall of as yet unknown origin exists just outside the event horizon that destroys the entanglement between infalling particles and outgoing radiation, effectively eliminating the information loss problem, albeit at a significant cost.
\end{enumerate}

The purpose of the present paper is to set up a semi-classical model that explicitly realizes an alternative proposal to resolve the information loss problem. This proposal is based on the principle that there are no true singularities in nature. This idea is not new. It was first put forward by Frolov and Vilkovisky\cite{Frolov1981,Frolov1979} and discussed more recently by \cite{Ashtekar2005}\footnote{See also Varadarajan \cite{Varadarajan2008}}.  We start with the assumption that in the correct theory the classical singularity will be replaced by a semi-classical non-singular region. Moreover, we assume that the dynamics of gravitational collapse and evaporation in this region can be described by effective semi-classical equations of motion. Our first goal is to construct a suitable ``quantum corrected''\footnote{These quantum corrections ultimately derive from the underlying, as yet unknown, quantum gravity theory.  Later on we will discuss corrections to the action which mimic Hawking radiation.  These corrections are derived from quantization of matter fields but, to avoid confusion, we will refer to them as radiation corrections or radiation terms.} lagrangian, potentially relevant to four dimensional black holes, from which to derive these semi-classical equations. Such an action must have three key attributes: 
\begin{enumerate}
\item It must incorporate modifications to Einstein gravity at short distances that resolve the singularity.
\item It must have a radiation back reaction term. 
\item At large distance scales it must accurately reproduce Einstein's theory.
\end{enumerate}

The expectation is that the resulting gravitational collapse and subsequent evaporation will produce neither an event horizon nor a singularity. Instead the result will be a completely non-singular spacetime that contains a closed dynamical horizon bounding a compact trapped region. At late times there will therefore be no impediment to information about the collapsed matter escaping to null infinity.
Such a scenario was discussed in \cite{Frolov1981,Frolov1979} and first made explicit
by Sean Hayward \cite{Hayward2006}.
More recently, a compact, dynamical trapping horizon was realized via the numerical simulation of the spherical collapse of a massless scalar field in \cite{Ziprick2009}. In this work the singularity resolution and energy loss were modelled by introducing explicit modifications to the gravitational potential in the equations of motion. Since the equations were not derived from a diffeomorphism invariant action, they did not preserve the constraint algebra and were non-conservative. A more systematic energy conserving model was later constructed \cite{Ziprick2010} and used to show that non-singular black holes could indeed be formed via gravitational collapse\footnote{See also \cite{Hossenfelder2010} and \cite{Zhang2014} for closely related work and \cite{Roman1983} for early work on singularity resolved black holes.} provided that the effective quantum corrections to the gravitational potential were introduced within the framework of a  (spherical symmetry preserving) diffeomorphism invariant action. The equations in this model were derived from a variational principle and hence are conservative, but they did not contain a mechanism for describing Hawking radiation.  
  
In the following, we continue the above program by constructing a set of dynamical equations for four D (and higher) gravity that lead to non-singular collapse, are energy conserving and include radiation back-reaction.
Modelling the effect of Hawking radiation is difficult and so far it is not well understood in more than two dimensions.  In two dimensions, however, the conformal anomaly can be calculated at one loop order and integrated to derive the non--local Polyakov action \cite{Polyakov1981},
\begin{align}
I_{Poly} & \sim -\int d^2x \sqrt{-g} \Ri \frac{1}{D^2} \Ri,
\label{eq:Polyakov0}
\end{align}
where $g$ is the determinant of the 2D metric, $g_{\mu \nu}$, $\Ri$ is the Ricci scalar calculated with $g_{\mu \nu}$ and $D$ is the covariant derivative compatible with $g_{\mu \nu}$.  This one loop effective action is exact in the large $N$ limit, where $N$ is the number of conformally coupled scalar fields.  It has been studied extensively in the context of 2D models for black hole evaporation \cite{Callan1992, Russo1992a, Ashtekar2011, Ashtekar2011a}. It has also been used in a toy model to simulate Hawking radiation in four dimensional, spherically symmetric Einstein gravity \cite{Ayal1997, Parentani1994}. This is the approach we will adopt. The new, and crucial features of our analysis are two-fold: first we will formulate the theory in terms of a local, diffeomorphism invariant effective action. Our equations are therefore energy conserving and allow for the use of slicings that extend into the horizon. Secondly, the action which is our starting point is a generalization of the spherically symmetric Einstein action that allows us to use quantum movitated corrections that resolve the singularity in the vacuum solution. Since the class of lagrangians we consider obey a Birkhoff theorem, this guarantees that gravitational collapse will yield an exterior spacetime that asymptotes to the corresponding non-singular solution. We therefore expect a complete absence of singularities in the dynamical evolution of the collapsing matter as found in \cite{Ziprick2010}. This will in turn permit us to examine via rigorous calculations, albeit in a simplistic model, the question posed above: How does singularity resolution affect the end point of gravitational collapse and Hawking radiation?  

Recently several other models have been proposed to investigate non--singular, radiating black holes \cite{Frolov2014,Bardeen2014,Torres2014,Mersini-Houghton2014} using different methods to model the radiation and singularity resolution than those used in this work.  Using the Polyakov term to model the radiation back reaction has the advantage that it is rigorous, at least in 2D.  It is also useful to have a variety of plausible models to hunt for generic features of the formation and evapouration process.

In the present paper, we describe the model and derive the equations of motion both in conformal gauge and in P-G type coordinates.  The organization is as follows: Section \ref{sec:GravAction} discusses the classical action (without radiating terms). Section \ref{sec:RadiationTerms} discusses the addition of the radiation corrections via the generalized Polyakov action in local form. Section \ref{sec:nullcoveq} derives the equations of motion in conformal gauge and discusses singularity resolution in the energy momentum tensor while Section \ref{sec:rtcoveq} presents the Hamiltonian analysis and the equations of motion in P-G type coordinates. Section \ref{sec:Conclusions} closes with conclusions and prospects.  The numerical simulations of these equations are relegated to a subsequent paper.

\section{Effective Gravitational Action}
\label{sec:GravAction}

We start from Einstein gravity in $n$ space-time dimensions minimally coupled to a massless scalar field, whose action is:
\be
I_{(n)} = \frac{1}{16\pi G^{(n)}}\int d^nx \sqrt{-\overline{g}} \left({\cal R}(\overline{g})
- 8\pi G^{(n)} (\overline{D}\psi)^2
\right)
\label{eq:n_dimensional action}
\ee
where $G^{(n)}$ is the higher dimensional gravitational constant, $\Ri(\bar{g})$ is the Ricci scalar calculated using the $n$-dimensional metric, $\bar{g}$, $\bar{D}$ is the derivative compatible with $\bar{g}$ and $\psi$ is a scalar field.  After imposing spherical symmetry, integrating out the angular variables and absorbing a factor of $\sqrt{8 \pi G^{(n)}}$ into the scalar field to make it dimensionless, this action takes the form
 \cite{Maeda2008, Maeda2011}:
\begin{align}
I_{(2)}&=\frac{1}{l^{n-2}}\int d^2x \sqrt{-g}\Big[ R^{n-2}\Ri \nonumber\\
 & + (n-2)(n-3)R^{(n-4)}(DR)^2 \nonumber \\
 & + (n-2)(n-3)R^{(n-4)} 
 -R^{n-2}(D\psi)^2\Big]
\label{eq:dim_reduced_action}
\end{align}
where $R$ is the areal radius, $\Ri$ is the Ricci scalar of $g$ which is the $(t,x)$ part of the higher dimensional metric: 
\be
ds^2 = g_{\mu\nu}dx^\mu dx^\nu + R^2 d\Omega^{(n-2)}
\ee 
and we have defined the length parameter:
\be
l^{(n-2)}:= \frac{16\pi G^{(n)}}{{\cal A}_{(n-2)}}
\ee
with $A_{(n-2)}$ being the invariant volume of a unit $n-2$-sphere.  In terms of these parameters the well known Schwarzschild-Tangherlini solution is:
\bea
ds^2 &=& -\left(1-\frac{l^{(n-2)}M}{(n-2) R^{n-3}}\right)dt^2 \nonumber\\
  & & + \left(1-\frac{l^{(n-2)}M}{(n-2) R^{n-3}}\right)^{-1}dR^2 + R^2 d\Omega^{(n-2)}
\label{eq:st}
\eea
where $M$ is the ADM mass in the vacuum case.

We now generalize the above allowing the coefficients of each of the three terms in \eqref{eq:dim_reduced_action} to be arbitrary functions of the areal radius $R(x)$, as,
\begin{align}
  \label{eq:Action2}
  I_{general}&=\frac{1}{l^{n-2}}\int d^2x \sqrt{-g} \times \nonumber \\
  & \Big\{ \phi(R)\Ri  + h(R)(DR)^2 + V(R) + B(R)(D\psi)^2 \Big\}.
\end{align}
The above is the most general action that contains at most two derivatives of the metric and areal radius and yields a Hamiltonian that is quadratic in their conjugate momenta.  It can in principle be further generalized to take the general form of a dimensionally reduced higher curvature Lovelock gravity, whose equations of motion are second order, but higher order in momenta. See \cite{Kunstatter2012,Kunstatter2013,Taves2013} for a Hamiltonian analysis of spherically symmetric Lovelock gravity. We relegate the investigation of such theories to future work.

 For \eqref{eq:Action2} to reduce to the GR case, two conditions must be satisfied,

\begin{equation}
\label{eq:condition1}
\phi = -B = R^{n-2}
\end{equation}
\begin{equation}
\label{eq:condition2}
h=V=\phi^{\prime \prime},
\end{equation}
where a prime means differentiation with respect to the areal radius.  In this work we are concerned with the case where \eqref{eq:condition2} is obeyed reducing the number of free functions by two (although, for the purpose of generality we do not substitute \eqref{eq:condition2} into our equations until we discuss important physical results).  As we will show \eqref{eq:condition2} will be necessary to remove the singularity in the vacuum solution while simultaneously removing another one in the radiating term.  It will also become important when defining a mass function (see appendix \ref{app:mass}) and when finding boundary terms at infinity which make the variational principal well defined (see appendix \ref{app:boundary}).

The action \eqref{eq:Action2} belongs to the class of theories called generic 2-D dilaton gravity  (see \cite{Grumiller2002} and \cite{Gegenberg2010} for reviews). These theories obey a Birkhoff theorem\cite{Louis-Martinez1994}. The most general vacuum solution can be found by defining a new metric $\tilde{g}_{\mu\nu} := \omega^2 g_{\mu\nu}$ which puts the action in the form 
\be
I = l^{-(n-2)} \int d^2x \sqrt{-\tilde{g}}\left( \phi {\cal R}(\tilde{g}) +V/\omega^2\right)
\ee
whose equations of motion can readily be solved \cite{Louis-Martinez1994}.  The corresponding metric in the current parameterization is:
\begin{align}
\label{eq:ds2vacuum}
ds^2 =& \frac{j}{\omega^{2}}\Bigg[-\left(1-\frac{l^{(n-2)}{M}}{j}\right)dt^2 \nonumber\\
& +  \left(1-\frac{l^{(n-2)}{M}}{j}\right)^{-1} \left(\frac{
\phi^\prime}{j}\right)^2dR^2\Bigg] +R^2 d\Omega^2
\end{align}
where 
\begin{align}
\ln(\omega^2)&:=\int \frac{h}{\phi^\prime}dR\\
j&:=\int ( \phi^\prime V/\omega^2 )dR.
\label{eq:2Dsoln}
\end{align}
The solution contains a single parameter, $M$, and has at least one Killing vector ($\partial/\partial t$). There are horizons whenever $j=l^{(n-2)}M$, and the Killing vector is timelike in the asymptotic region, $j>l^{(n-2)}M$. 
Most importantly, the arbitrary functions of the areal radius $\phi(R)$, $h(R)$ and $V(R)$  can, as we shall see, be chosen so that the resulting vacuum solutions are non-singular.


As an example of a non-singular metric that can be obtained as a solution to the equations derived from an action of the form \eqref{eq:Action2} consider the following metric that was originally proposed by Poisson and Israel \cite{Poisson1990}:
\bea
ds^2 &=& -\left(1-\frac{l^{2}MR^2 }{2(R^3+\nu^3)}\right)dt^2\nonumber\\
& &+\left(1-\frac{l^{2}MR^2 }{2(R^3+\nu^3)}\right)^{-1}{dR^2} +R^2 d\Omega^2. \nonumber\\
\label{eq:ZKBH}
\eea
This is of the same form as \eqref{eq:ds2vacuum} with the following identifications: 
\begin{align}
\label{eq:phiprimeZip}
\phi^\prime &=j=2\frac{R^3+\nu^3}{R^2}\\
\label{eq:phiZip}
\phi&= \frac{R^4-2\nu^3R}{R^2}\\
\label{eq:omegaZip}
\omega^2 &= j =2\frac{R^3+\nu^3}{R^2} \\
\label{eq:hZip}
h &= \ln(\omega^2)^\prime \phi^\prime =j^\prime = 2\frac{R^3-2\nu^3}{R^3}\\
\label{eq:VZip}
V&= \omega^2 j^\prime/\phi^\prime = 2\frac{R^3- 2\nu^3}{R^3}
\end{align}
In the above $\nu$ is the parameter that determines the scale of the quantum corrections. The above metric approaches a deSitter metric as $R\to0$ and hence is manifestly non-singular. The resulting spacetime has two horizons and the effective stress energy tensor violates the weak energy condition at short length scales (of the order of $\nu$). The properties of this metric are discussed more fully in \cite{Ziprick2010}.
They can easily be generalized to higher dimensions with:
\be
j(R) = (n-2) \frac{R^{n-1}+\nu^{n-1}}{R^2}
\ee

A qualitatively different example of a non-singular static spacetime was derived from polymer quantum gravity in \cite{Modesto2004} and \cite{Peltola2009}.  This spacetime contains a single bifurcative horizon that surrounds an infinite Kasner type universe on the interior.  It describes, in effect, a wormhole whose minimum throat radius shrinks to the polymerization scale before re-expanding to infinity. Details of how it fits into the current formalism are given in Appendix \ref{app:KPBH}.

\section{Radiation Terms: The Generalized Polyakov Action}
\label{sec:RadiationTerms}

The radiation term \eqref{eq:Polyakov0} assumes that the matter under consideration is minimally coupled to gravity, ie $B$ is a constant in \eqref{eq:Action2}.  We will use a more general version that allows for more general matter couplings \cite{Nojiri1998, Kummer1998, Medved1999, Medved2001},

\begin{align}
I_{Poly} & \sim -\int d^2x \sqrt{-g} \Bigg[ \Ri \frac{1}{D^2} \Ri \nonumber \\ 
& + b(R)\left( \frac{1}{D^2}\Ri - \ln \mu^2  \right) (DR)^2 +c(R)\Ri \Bigg],
\label{eq:Polyakov1}
\end{align}
where $\mu$ is a constant related to the renormalization procedure used to obtain the Polyakov action.  The forms of $b$ and $c$ have been suggested by \cite{Medved2001} to be $b=-3(B^\prime)^2/B^2$ and $c=-6\ln{B}$.  Since it may be interesting to also investigate the case where $b=c=0$ (as done by \cite{Ayal1997}) we perform the algebra without making any assumptions about $b$ and $c$ until we discuss important, physical results (see appendix \ref{app:boundary}).

The non-local character of the actions \eqref{eq:Polyakov0} and \eqref{eq:Polyakov1} is often dealt with by working in double--null co--ordinates, 
\be
ds^2=e^{2f}dudv,
\ee
 where $f$ is a scalar function and the Ricci scalar takes the form

\begin{equation}
  \label{eq:Riccinull1}
  \Ri=-8e^{-2f} f_{,uv} = 2D^2f.
\end{equation}
It is sometimes advantageous, however, to work in other co--ordinate systems, particularly when performing numerical simulations that go past horizon formation (see numerics section of \cite{Garfinkle1995}).  In this case auxiliary fields, $z_1$ and $z_2$ can be used to write the action as \cite{Hayward1995, Gegenberg2012, Medved2001}

\begin{align}
  \label{eq:Polyakov2}
  I_{Poly}& \sim - \int \sqrt{-g} [\Ri(z_1+z_2)
 + D_Az_2D^Az_1 \nonumber \\
& + b (DR)^2(z_1-\ln \mu^2) + c \Ri ],
\end{align}
where $A=0,1$.  The equations of motion for $z_1$ and $z_2$ are given by

\begin{equation}
  \label{eq:z1EoM}
  D^2z_1 - \Ri = 0,
\end{equation}
\begin{equation}
  \label{eq:z2EoM}
  D^2z_2 - \Ri - b(DR)^2 = 0.
\end{equation}
Inserting \eqref{eq:z1EoM} and \eqref{eq:z2EoM} into \eqref{eq:Polyakov2} gives \eqref{eq:Polyakov1}.

Inspired by \eqref{eq:Action2} and \eqref{eq:Polyakov2} we define the action that we will consider for the rest of this work as
\begin{align}
  \label{eq:Action2b}
  I&=\frac{1}{l^{n-2}}\int d^2x \sqrt{-g} \Big\{ \phi(R)\Ri  + h(R)(DR)^2 + V(R) \nonumber \\
&+ W\big[\Ri(z_1+z_2)
 + D_Az_2D^Az_1 \nonumber \\
& + b(R) (DR)^2(z_1-\ln \mu^2) + c(R) \Ri \big] 
+ B(R)(D\psi)^2 \Big\}
\end{align}
where $W$ is a coupling constant.




It is worth noting here that Ayal and Piran \cite{Ayal1997} (see also \cite{Parentani1994}) considered a model similar to this one.  In fact they considered the equations of motion, in null gauge, that would come from varying \eqref{eq:Action2b} with the fields given as those for the 4D GR case with $b=c=0$ and $z_1=z_2$.  They eventually modified those equations of motion to remove a singularity at $R=0$ in the effective stress energy tensor.  The subsequent equations did not obey the Bianchi identities, however, since they altered the equations of motion directly and not their corresponding action.


\section{Double Null Co--ordinates}
\label{sec:nullcoveq}

In this section we  derive the covariant equations of motion, constraints and effective energy momentum tensor from the action \eqref{eq:Action2b} in null gauge. 

\subsection{Action}
 We start with the metric

\begin{equation}
  \label{eq:nullg1}
  g_{\mu \nu} = 
\begin{bmatrix} 
   0    & e^{2f}/2 \\ 
  e^{2f}/2 & 0
\end{bmatrix}
\end{equation}
from which the Ricci scalar is given by \eqref{eq:Riccinull1}.  Using \eqref{eq:nullg1} the action, \eqref{eq:Action2b} becomes

\begin{align}
  \label{eq:Action2null}
  I&=\int d^2x \Big\{-4\phi(R)f_{,uv}  + 2h(R)R_{,u}R_{,v} + \frac{1}{2}e^{2f}V(R) \nonumber \\
&+ W\bigg[-4f_{,uv} (z_1 + z_2) + (z_{1,u}z_{2,v}+z_{2,u}z_{1,v}) \nonumber \\
&+ 2b(R)R_{,u}R_{,v}(z_1-\ln \mu^2)  -4c(R)f_{,uv} \bigg] \nonumber \\ 
&+ 2B(R)\psi_{,u}\psi_{,v} \Big\}
\end{align}
where we have absorbed the factor of $l^{n-2}$ into the action. 

\subsection{Equations of Motion}
 Varying \eqref{eq:Action2null} gives the equations of motion

\begin{align}
  \label{eq:REoM2b}
& \frac{\delta I}{\delta R} = 0 = \nonumber \\
&-4\phi^{\prime}f_{,uv} -2 h^{\prime} R_{,u}R_{,v} -4hR_{,uv} + (e^{2f}/2)V^{\prime} \nonumber \\
& + 2B^{\prime}\psi_{,u}\psi_{,v} +W\bigg[ -2(b^\prime R_{,u} R_{,v} + 2bR_{,uv})(z_1 -\ln \mu^2)
\nonumber \\ 
& -2b(R_{,u}z_{1,v} + R_{,v}z_{1,u}) 
-4c^\prime f_{,uv} \bigg],
\end{align}

\begin{align}
  \label{eq:fEoM2b}
&\frac{\delta I}{\delta f} = 0 = \nonumber \\
&-4R_{,uv}\phi^{\prime}-4R_{,u}R_{,v}\phi^{\prime \prime} + e^{2f}V \nonumber \\
&+W \bigg[-4z_{1,uv} -4z_{2,uv} 
-4R_{,uv}c^{\prime}-4R_{,u}R_{,v}c^{\prime \prime}\bigg],
\end{align}

\begin{align}
  \label{eq:psiEoM2b}
  \frac{\delta I}{\delta \psi} = 0 = -2B^\prime R_{,u}\psi_{,v} -4B\psi_{,uv} -2B^\prime R_{,v}\psi_{,u},
\end{align}

\begin{align}
  \label{eq:z1EoM2b}
\frac{\delta I}{\delta z_2} = 0 = -4Wf_{,uv} -2W z_{1,uv}
\end{align}
and

\begin{align}
  \label{eq:z2EoM2b}
\frac{\delta I}{\delta z_1} = 0 = -4Wf_{,uv} -2Wz_{2,uv} +2WbR_{,u}R_{,v}.
\end{align}
Equations \eqref{eq:z1EoM2b} and \eqref{eq:z2EoM2b} can be used to write \eqref{eq:REoM2b} and \eqref{eq:fEoM2b} with out any dependence on $z_1$ or $z_2$ as

\begin{align}
  \label{eq:REoM2c}
&-4\phi^{\prime}f_{,uv} -2 h^{\prime} R_{,u}R_{,v} -4hR_{,uv} + (e^{2f}/2)V^{\prime} + 2B^{\prime}\psi_{,u}\psi_{,v} \nonumber \\
&+W\bigg[ -2(b^\prime R_{,u} R_{,v} + 2bR_{,uv})(-2f -\ln \mu^2)
\nonumber \\ 
& +4b(R_{,u}f_{,v} + R_{,v}f_{,u}) 
-4c^\prime f_{,uv} \bigg] = 0
\end{align}
and

\begin{align}
  \label{eq:fEoM2c}
&-4R_{,uv}\phi^{\prime}-4R_{,u}R_{,v}\phi^{\prime \prime} + e^{2f}V \nonumber \\
&+W \bigg[16f_{,uv} -4bR_{,u}R_{,v} 
-4R_{,uv}c^{\prime}-4R_{,u}R_{,v}c^{\prime \prime}\bigg] = 0
\end{align}
Using the GR values for $\phi$, $h$, $V$ and $B$ as well as $W=-\alpha/8\pi$, $b=c=0$, $n=4$ and $u \to -u$ (due to different sign conventions) we find that
\eqref{eq:REoM2c} and \eqref{eq:fEoM2c} reduce to (11a) and (11b) of \cite{Ayal1997} (where $\alpha$ is the coupling constant used in that work).

\subsection{Constraint Equations}
To find the constraint equations we use the most general metric

\begin{equation}
  \label{eq:generalg1}
  g_{\mu \nu} = 
\begin{bmatrix} 
   A    &  e^{2f}/2 \\ 
  e^{2f}/2  & C
\end{bmatrix}
\end{equation}
which corresponds to the Ricci scalar

\begin{align}
  \label{eq:generalRicci1}
  \Ri & = (-g)^{-2} \times \nonumber \\
 \Bigg\{ & A\left[(C_{,u})^2/2 + C_{,v}A_{,v}/2 - e^{-2f}C_{,v}f_{,u} \right] \nonumber \\
       + & C\left[(A_{,v})^2/2 + C_{,u}A_{,u}/2 - e^{-2f}A_{,u}f_{,v} \right] \nonumber \\
 + & (e^{2f}/2) [ -e^{2f}A_{,v}f_{,v}-e^{2f}C_{,u}f_{,u} \nonumber \\ 
+ & A_{,u}C_{,v}/2-C_{,u}A_{,v}/2+8ACf_{,u}f_{,v} ] \Bigg\} \nonumber \\
+ & (-g)^{-1}\Bigg\{A_{,vv} - 2e^{2f}f_{,uv} + C_{,uu} \Bigg\}
\end{align}
From \eqref{eq:generalRicci1} we can see the following useful relationships (up to boundary terms)

\begin{align}
  \label{eq:Avariation1}
  & \Bigg(\frac{\delta}{\delta A}\int d^2x \sqrt{-g} \phi \Ri \Bigg\|_{A=C=0} = \nonumber \\
& \left(2\phi_{,vv} -4f_{,v}\phi_{,v}  \right)e^{-2f}= \nonumber \\
& \left(2\phi^{\prime \prime} (R_{,v})^2+2\phi^{\prime} R_{,vv}  -4f_{,v}\phi^{\prime}R_{,v}  \right)e^{-2f}
\end{align}

\begin{align}
  \label{eq:Avariation2}
  & \Bigg(\frac{\delta}{\delta A}\int d^2x \sqrt{-g} h (DR)^2 \Bigg\|_{A=C=0} = \nonumber \\
& - 2h e^{-2f} (R_{,v})^2
\end{align}
This gives the constraint

\begin{align}
  \label{eq:constraint1}
& \frac{\delta I}{\delta A}\Bigg\|_{A=C=0} = 0 = \nonumber \\
  & 2e^{-2f} [\phi^{\prime \prime} (R_{,v})^2 + \phi^\prime R_{,vv} - 2\phi^\prime f_{,v}R_{,v}] - 2e^{-2f}h(R_{,v})^2 \nonumber \\
& -2e^{-2f}B(\psi_{,v})^2 \nonumber \\
& + W \{ 2e^{-2f} [z_{1,vv} + z_{2,vv} -2f_{,v}z_{1,v} -2f_{,v}z_{2,v}  ] \nonumber \\
& - 2e^{-2f}z_{1,v}z_{2,v}
- 2e^{-2f}b(z_1-\ln \mu^2)(R_{,v})^2 \nonumber \\
& + 2e^{-2f} [c^{\prime \prime} (R_{,v})^2 + c^\prime R_{,vv} - 2c^\prime f_{,v}R_{,v}] \}
\end{align}
Note from \eqref{eq:z1EoM2b} that $z_1(u,v)=-2f(u,v)+p_u(u)+p_v(v)$, where $p_u$ and $p_v$ are some functions.  If we assume that $p_u$ and $p_v$ are constants and re-arrange \eqref{eq:constraint1} and use \eqref{eq:z1EoM2b} and \eqref{eq:z2EoM2b} we get

\begin{align}
  \label{eq:constraint3}
&-R_{,vv} + 2f_{,v}R_{,v} = 
\frac{\phit^{\prime \prime}+2Wb(f+\ln|\mu|)-h}{\phit^\prime } (R_{,v})^2 \nonumber \\
& - \frac{B}{\phit^\prime }(\psi_{,v})^2  - \frac{4W}{\phit^{\prime 2}} \{f_{,vv} - (f_{,v})^2 -{\bar b}_{,v}/4 \}
\end{align}
where we define

\begin{equation}
  \label{eq:phit1}
  \phit := \phi + Wc
\end{equation}
\begin{equation}
  \label{eq:bu1}
  {\bar b}(v):= \int du (bR_{,u}R_{,v})
\end{equation}
and we have used \eqref{eq:z2EoM2b} to write

\begin{equation}
  \label{eq:z2vv1}
  z_{2,vv} = -2f_{,vv} + {\bar b}_{,v}.
\end{equation}
There is also a second constraint equation which can be found by swapping $u$ and $v$ in \eqref{eq:constraint3}.
\subsection{Effective Stress Energy Tensor}
\label{sec:EMTensor}
We now use the equations of motion and Einstein's equations to calculate the energy momentum tensor.  Using (2.18) of \cite{Maeda2011} we find that the non-angular components of the Einstein tensor in $n$ dimensions are given by

\begin{align}
  \label{eq:Einsteinuunull1}
  & G_{uu}^{(n)} = \frac{(n-2)}{R}   (-R_{,uu}+2f_{,u}R_{,u})
\end{align}
\begin{align}
  \label{eq:Einsteinvvnull1}
  & G_{vv}^{(n)} = \frac{(n-2)}{R}   (-R_{,vv}+2f_{,v}R_{,v})
\end{align}
\begin{align}
  \label{eq:Einsteinuvnull1}
  & G_{uv}^{(n)} = \frac{(n-2)}{R}    
\left[ R_{,uv} -\frac{(n-3)}{R}\left(\frac{e^{2f}}{4}-R_{,u}R_{,v}\right) \right]
\end{align}
Concentrating on the $u-v$ component, using the equation of motion \eqref{eq:fEoM2c} to solve for $R_{,uv}$ and the expression for the mass \eqref{eq:mass1}

\begin{align}
(D R)^2 &= \frac{j\omega^2}{\phi^{\prime\,2}}\left(1-\frac{M}{j}\right)\\
        &= 4 e^{-2f} R_{,u}R_{,v}
\end{align}
(where a factor of $l^{n-2}$ has been absorbed into $M$) gives the following expression for the effective stress tensor
\begin{align}
\label{eq:Einsteinnull2}
G^{(n)}_{uv}&= \frac{(n-2)}{\phit^\prime R}\Bigg\{
\frac{e^{2f}}{4} \frac{M}{j} \left(\phit'' - \frac{(n-3)\phit'}{R}\right) \nonumber \\
&+ W\Bigg[ 4f_{,uv}-\frac{e^{2f}}{4}\left(c^{\prime \prime} +b\left(1-\frac{M}{j}\right)\right)\Bigg] \Bigg\}.
\end{align}
In the above we have assumed only that \eqref{eq:condition2} is satisfied, which also implies that  $j=\phi' = \omega^2$.  This is necessary in order to eliminate terms in the stress energy tensor that are singular at $R=0$.  These conditions are satisfied for the two horizon quantum corrected black hole described by \eqref{eq:phiprimeZip}-- \eqref{eq:VZip}. Assuming reasonable regularity conditions on the mass function and $c$ and $b$ at the origin, the singularity observed by Piran and Ayal in the effective stress tensor is resolved as long as the singularity in $1/j$ is resolved.  Similar statements apply to the diagonal components of the energy momentum tensor as we can see by comparing \eqref{eq:Einsteinvvnull1} to \eqref{eq:constraint3}.

\section{Hamilton's Equations in Non--Null Co--ordinates}
\label{sec:rtcoveq}

To find the equations of motion in non-null co--ordinates.  We start with the general, ADM metric \cite{Arnowitt2004},

\begin{equation}
  \label{eq:generalds2}
  ds^2 = -N^2dt^2 + \Lambda^2(N_rdt + dx)^2.
\end{equation}
For later use, we compute the following quantities:

\begin{align}
&(DR)^2 =-{R_y}^2+\Lambda^{-2}{R_{,x}}^2,\label{defF}\\
\sqrt{-g_{(2)}}D^2 R=&-\partial_t(\Lambda R_{,y})+\partial_x(\Lambda N_r R_{,y}+\Lambda^{-1}NR_{,x}),\label{DDR}
\end{align} 
where we define the operator, $,y$ acting on some field, $\beta$ by 

\begin{equation}
\beta_{,y}:=N^{-1}( \beta_{,t}-N_r\beta_{,x}).\label{defy}
\end{equation}
Using the metric \eqref{eq:generalds2} the Ricci scalar can be written as

\begin{align}
  \label{eq:generalRicciy1}
  \sqrt{-g} \Ri & = 2\{N_{r,x}N_{,y}\Lambda N^{-1} -2N_{r,x} \Lambda_{,y} - \Lambda (N_{r,x})_{,y} \nonumber \\
&+ N^{-1} \Lambda (N_{r,x})^2 + N \Lambda_{,yy} - (N_{,x} \Lambda^{-1})_{,x}\}
\end{align}

To write the action in a form that  lends itself to Hamiltonian analysis we define

\begin{align}
  \label{eq:Phi1}
  \Phi(R(t,x),t,x)& := \phi(R) + W(z_1 + z_2) + Wc(R) \nonumber \\
  & =\phit(R) + W(z_1 + z_2)
\end{align}
and

\begin{equation}
  \label{eq:h21}
  \hz(R(t,x),z_1(t,x)):=h(R) +Wb(R)(z_1 - \ln \mu^2)
\end{equation}
With these definitions the action looks like

\begin{align}
  \label{eq:Action3a}
  I&=\frac{1}{l^{n-2}}\int d^2x \sqrt{-g} \times \nonumber \\
& \Big\{ \Phi\Ri  + \hz(DR)^2 + V + WD_Az_2D^Az_1 + B(D\psi)^2 \Big\}
\end{align}
We first work with the term containing the Ricci scalar.  From \eqref{eq:generalRicciy1} we can calculate this term up to total derivatives (t.d.) to be

\begin{align}
   \label{eq:generalRicciphit2}
  \sqrt{-g} \Ri\Phi=2\Phi_{,y}(N_{r,x}\Lambda- N\Lambda_{,y}) -2N(\Phi_{,x} \Lambda^{-1})_{,x} +t.d.
\end{align}
Using this the action \eqref{eq:Action3a} becomes

\begin{align}
  \label{eq:Action3}
  I&=\frac{1}{l^{n-2}}\int d^2x \times \nonumber \\
& \Big\{2N^{-1}(\Phi_{,t} -N_{r}\Phi_{,x})(N_{r,x}\Lambda-(\Lambda_{,t}- N_{r} \Lambda_{,x})) \nonumber \\
& -2N(\Phi_{,x} \Lambda^{-1})_{,x} +N\Lambda V \nonumber \\
&+ \hz [-N^{-1}\Lambda R_{,t}^{\phantom {,x}2} +2N_rN^{-1}\Lambda R_{,t} R_{,x} \nonumber \\
&+(N\Lambda^{-1} - N_r^2N^{-1}\Lambda) R_{,x}^{\phantom {,x}2} ] \nonumber \\
& + W [-N^{-1}\Lambda z_{1,t}z_{2,t} + N_rN^{-1}\Lambda(z_{1,x}z_{2,t} + z_{1,t}z_{2,x}) \nonumber \\
&+(N\Lambda^{-1} - N_r^2N^{-1}\Lambda)z_{1,x}z_{2,x} ] \nonumber \\
&+ B [-N^{-1}\Lambda \psi_{,t}^{\phantom {,x}2} +2N_rN^{-1}\Lambda\psi_{,t}\psi_{,x} \nonumber \\
&+(N\Lambda^{-1} - N_r^2N^{-1}\Lambda)\psi_{,x}^{\phantom {,x}2} ] \Big\}
\end{align}
We can see from \eqref{eq:Action3} that the conjugate momenta corresponding to $N$ and $N_r$ are zero, $P_N = P_{N_r} = 0$.  The remaining conjugate momenta, corresponding to $\Lambda$, $R$, $z_1$, $z_2$ and $\psi$ are given by

\begin{align}
  \label{eq:PLambda1}
  \PL& = -l^{-(n-2)}2N^{-1}(\Phi_{,t} - N_r \Phi_{,x}) \nonumber \\
& =  -l^{-(n-2)}2(\phit^\prime R_{,y} +W(z_1 + z_2)_{,y}),
\end{align}

\begin{align}
  \label{eq:PR1}
  P_R =& l^{-(n-2)}[2N^{-1}\phit^\prime(N_{r,x}\Lambda-(\Lambda_{,t}- N_{r} \Lambda_{,x})) \nonumber \\
& -2\Lambda \hz N^{-1}(R_{,t} - N_rR_{,x})] \nonumber \\
=& l^{-(n-2)}[2N^{-1}\phit^\prime N_{r,x}\Lambda - 2\phit^\prime \Lambda_{,y} -2\Lambda \hz R_{,y}],
\end{align}

\begin{align}
  \label{eq:Pz11}
  P_{z1} =& l^{-(n-2)}[2N^{-1}W(N_{r,x}\Lambda-(\Lambda_{,t}- N_{r} \Lambda_{,x})) \nonumber \\
& +W(-N^{-1}\Lambda z_{2,t} +N_r N^{-1}\Lambda z_{2,x})] \nonumber \\
=& l^{-(n-2)}W[2N^{-1} N_{r,x}\Lambda- 2\Lambda_{,y} -\Lambda z_{2,y}], 
\end{align}

\begin{align}
  \label{eq:Pz21}
  P_{z2} =& l^{-(n-2)}[2N^{-1}W(N_{r,x}\Lambda-(\Lambda_{,t}- N_{r} \Lambda_{,x})) \nonumber \\
& +W(-N^{-1}\Lambda z_{1,t} +N_r N^{-1} \Lambda z_{1,x})] \nonumber \\
=& l^{-(n-2)}W[2N^{-1} N_{r,x}\Lambda- 2\Lambda_{,y} -\Lambda z_{1,y}],
\end{align}
and
\begin{align}
  \label{eq:Ppsi1}
  P_{\psi}& = l^{-(n-2)}B[-2N^{-1}\Lambda \psi_{,t} +2N_r N^{-1}\Lambda \psi_{,x}] \nonumber \\
& =-2l^{-(n-2)}B\Lambda \psi_{,y} 
\end{align}
respectively.  By combining \eqref{eq:PLambda1}, \eqref{eq:PR1}, \eqref{eq:Pz11} and \eqref{eq:Pz21} we get

\begin{align}
  \label{eq:z1y1}
  z_{1,y}& = \frac{-\Lambda\PL/2 -\Lambda\phit^\prime R_{,y} + P_{z1} -P_{z2}}{2W\Lambda}\nonumber \\
  & = -\Sd -\frac{P_{z2}}{W\Lambda},
\end{align}

\begin{align}
  \label{eq:z2y1}
  z_{2,y}& = \frac{-\Lambda\PL/2 -\Lambda\phit^\prime R_{,y} - P_{z1} +P_{z2}}{2W\Lambda}\nonumber \\
  & = -\Sd -\frac{P_{z1}}{W\Lambda}
\end{align}
and
\begin{align}
  \label{eq:Ry2}
  R_{,y}& = \frac{\phit^{\prime}(-\Lambda\PL/2+P_{z1}+P_{z2}) -2WP_R}{\Lambda(\phit^{\prime 2} + 4W \hz)} \nonumber \\
  & = \frac{-\phit^\prime \Sa -2W P_R}{\Sb}, 
\end{align}
where, for ease of notation we have defined

\begin{align}
  \label{eq:S1}
  \Sa:=\Lambda\PL/2 -P_{z1} -P_{z2},
\end{align}

\begin{align}
  \label{eq:S2}
  \Sb:=\Lambda(\phit^{\prime 2} +4W \hz)
\end{align}
and
\begin{align}
  \label{eq:S3}
  \Sd:= \frac{2\hz \Sa - \phit^\prime P_R}{\Sb}.
\end{align}
and we have absorbed a factor of $l^{n-2}$ into all of the conjugate momenta.  Plugging \eqref{eq:z1y1}, \eqref{eq:z2y1}, \eqref{eq:Ry2} as well as \eqref{eq:PR1} and \eqref{eq:Ppsi1} into the action \eqref{eq:Action3} it can be shown that the Hamiltonian density, ${\cal H}$, can be written as a sum of two constraints,

\begin{align}
  \label{eq:Hamconstraint1}
  {\cal H} = NH + N_rH_r + t.d.,
\end{align}
where the constraints are given by
\begin{align}
\label{eq:H2b}
H=& 
2\left(\frac{\phit^{\prime}R_{,x}}{\Lambda}\right)_{,x}
 -\frac{\hz R_{,x}^{\phantom {,x}2}}{\Lambda}
-\Lambda V \nonumber \\
&+W\left[-\frac{z_{1,x} z_{2,x}}{\Lambda} +2\left(\frac{z_{1,x} +z_{2,x}}{\Lambda}\right)_{,x}\right] \nonumber \\
&+\frac{h_z\Sa^2 -\phit^\prime P_R\Sa -WP_R^2}{\Sb} -\frac{P_{z1}P_{z2}}{W\Lambda} \nonumber \\
&-\frac{P_{\psi}^2}{4\Lambda B} -\frac{B\psi_{,x}^{\phantom {,x}2}}{\Lambda}
\end{align}
and
\begin{align}
\label{eq:Hr2b}
H_r=-P_{\Lambda,x} \Lambda +P_R R_{,x} + P_{z1}z_{1,x} +P_{z2}z_{2,x} +P_{\psi}\psi_{,x}.
\end{align}
Note that we absorbed a factor of $l^{n-2}$ into $H^{(M)}$, $H_r^{(M)}$, $H^{(G)}$ and $H_r^{(G)}$ (in addition to all of the conjugate momenta).  

From \eqref{eq:Hamconstraint1} we can write down the equations of motion as

\begin{align}
  \label{eq:REoM1}
  R_{,t}=N\Sc +N_r R_{,x}
\end{align}

\begin{align}
  \label{eq:EoMPR1}
  P_{R,t}=&-\left(\frac{2N_{,x}}{\Lambda}\right)_{,x} \phit^\prime 
-\left(\frac{2N \hz R_{,x}}{\Lambda}\right)_{,x} \nonumber \\
& + \frac{N \hz^\prime {R_{,x}}^2}{\Lambda}  +N\Lambda V^\prime \nonumber \\
&  +N\frac{-2\phit^{\prime \prime}\Sa \Sc -\Lambda[2\phit^\prime \phit^{\prime \prime} +4W \hz^\prime] {\Sc}^2}{4W} \nonumber \\
& -\frac{NP_\psi^2 B^\prime}{4\Lambda B^2} +\frac{NB^\prime{\psi_{,x}}^2}{\Lambda}
+(N_r P_R)_{,x}
\end{align}

\begin{align}
  \label{eq:LambdaEoM1}
  \Lambda_{,t}=N\Lambda\frac{\Sd}{2} +(N_r \Lambda)_{,x}
\end{align}

\begin{align}
  \label{eq:PLEoM1}
  P_{\Lambda,t}=&-\frac{2N}{\Lambda}\left(\frac{\phit^\prime R_{,x}}{\Lambda} \right)_{,x}
-\frac{2N_{,x}\phit^\prime R_{,x}}{\Lambda^2}
+2NV \nonumber \\
&-2W\left[\frac{N}{\Lambda}\left(\frac{z_{1,x} +z_{2,x}}{\Lambda} \right)_{,x}  +N_{,x}\left(\frac{z_{1,x} +z_{2,x}}{\Lambda^2} \right) \right] \nonumber \\
&-\frac{N\PL \Sd}{2} 
+ \frac{NH}{\Lambda} +P_{\Lambda,x} N_r
\end{align}

\begin{align}
\label{eq:psiEoM1}
\psi_{,t}=& -\frac{NP_{\psi}}{2\Lambda B} +N_r \psi_{,x}
\end{align}

\begin{align}
\label{eq:PpsiEoM1}
P_{\psi,t}=& \Bigg(-\frac{2NB\psi_{,x}}{\Lambda}
+ N_r P_\psi \Bigg)_{,x}
\end{align}

\begin{align}
  \label{eq:z1EoM1}
  z_{1,t} = -N \Sd -\frac{NP_{z2}}{W\Lambda}
+N_r z_{1,x}
\end{align}

\begin{align}
  \label{eq:Pz1EoM1}
  P_{z1,t} =& \Bigg[-\frac{W(Nz_{2,x} +2N_{,x})}{\Lambda} +N_r P_{z1} \Bigg]_{,x}
   \nonumber \\
& -N\Lambda Wb\Bigg[{\Sc}^2 -\frac{{R_{,x}}^2}{\Lambda^2} \Bigg]
\end{align}

\begin{align}
  \label{eq:z2EoM1}
  z_{2,t} = -N\Sd -\frac{NP_{z1}}{W\Lambda}
+N_r z_{2,x}
\end{align}

\begin{align}
  \label{eq:Pz2EoM1}
  P_{z2,t} = \Bigg[-\frac{W(Nz_{1,x} +2N_{,x})}{\Lambda} +N_r P_{z2} \Bigg]_{,x}
\end{align}
where we used the following useful identity,
\begin{align}
  \label{eq:identity1}
  \frac{\Sa^2/\Lambda -\Sb {R_{,y}}^2}{4W}=\frac{\hz \Sa^2 - \phit^\prime P_R \Sa - W {P_R}^2}{\Sb}
\end{align}

Since our Hamiltonian is the sum of two first class constraints we must now pick two gauge choices.  We will pick 

\begin{align}
  \label{eq:chi1}
  \chi:=R-x =0.
\end{align}
which makes the spatial co--ordinate the areal radius.

Before choosing the second gauge notice that the mass function (see Appendix \ref{app:mass}) is given by

\begin{align}
M =& j\left(1- \frac{(\phi^\prime)^2}{j\omega^2}(D R)^2\right)\nonumber\\
 =& j\left(1- \frac{(\phi^\prime)^2}{j\omega^2}\left[\left(\frac{R_{,x}}{\Lambda}\right)^2 -\left(\frac{-\phit^\prime \Sa -2W P_R}{\Sb}\right)^2\right]\right)
\label{eq:generalM1}
\end{align}
where we used \eqref{eq:Ry2}.  This inspires the choice of gauge 

\be
\Lambda - \sqrt{\frac{(\phi^\prime R_{,x})^2}{j\omega^2}} =0.
\label{eq:Lambdagaugegeneral}
\ee
We are interested in the case where \eqref{eq:condition2} is satisfied (which implies $(\phit^\prime R_{,x})^2j^{-1}\omega^{-2}=1$) 
For this reason we choose
\begin{align}
  \label{eq:xi1}
  \xi:=\Lambda - 1 = 0.
\end{align}

For both the GR case and the case of \eqref{eq:phiprimeZip} - \eqref{eq:VZip} the mass function is well defined and is the boundary term which must be added to the Hamiltonian to make the variational principle well defined (see Appendix \ref{app:boundary}).  For our gauge choices (and assuming \eqref{eq:condition2}) the mass function is given by

\begin{align}
\label{eq:massPG}
M=& j\left(\frac{-\phit^\prime \Sa -2W P_R}{\Sb}\right)^2.
\end{align}
By comparing \eqref{eq:chi1} and \eqref{eq:xi1} to \eqref{eq:generalds2} we can also see that this gauge choice is regular and spatially flat at horizon formation, i.e. when 

\begin{align}
  \label{eq:Horizon1}
  (DR)^2 =0.
\end{align}

The consistency conditions on these two gauge choices can be obtained by setting ${\dot R}=0$ and ${\dot \Lambda}=0$ in \eqref{eq:REoM1} and \eqref{eq:LambdaEoM1}.  Note that we now use a dot to represent differentiation with respect to our time, $T$, in this choice of gauge and that a prime still represents differentiation with respect to the areal radius, $R$, which is now our spatial co--ordinate.  The consistency conditions are given by

\begin{align}
  \label{eq:gaugechi1}
  \frac{N_r}{N} = -\Sc
\end{align}
and

\begin{align}
  \label{eq:gaugexi1}
  \left(-N \Sc\right)^\prime +N\frac{\Sd}{2} = 0,
\end{align}
from which we can write $N_r$ as

\begin{align}
  \label{eq:consistency12}
  N_r=-\exp\left[\int dR \frac{\Sd}{2 \Sc} \right]
\end{align}
where we used \eqref{eq:chi1} and \eqref{eq:xi1} to redefine

\begin{align}
  \label{eq:S12}
  \Sa:=\PL/2 -P_{z1} -P_{z2},
\end{align}

\begin{align}
  \label{eq:S22}
  \Sb:=\phit^{\prime 2} +4W \hz,
\end{align}

\begin{align}
  \label{eq:S32}
  \Sd:= \frac{2\hz \Sa - \phit^\prime P_R}{\Sb}
\end{align}
and
\begin{align}
  \label{eq:Ry3}
  \Sc = \frac{-\phit^\prime \Sa -2W P_R}{\Sb}. 
\end{align}

$R$ and $\Lambda$ are no longer phase space variables and neither are their conjugate momenta.   $P_R$ and $P_\Lambda$ must be written in terms of the remaining phase space variables by setting the Hamiltonian and momentum constraints to zero.  Ie, we must solve

\begin{align}
\label{eq:H2c}
H=&0=
2\phit^{\prime \prime}
 -\hz
- V 
+W\left[-z_{1}^\prime z_{2}^\prime +2(z_{1}^{\prime \prime} +z_{2}^{\prime \prime})\right] \nonumber \\
&+\frac{h_z\Sa^2 -\phit^\prime P_R\Sa -WP_R^2}{\Sb} -\frac{P_{z1}P_{z2}}{W} -\frac{P_{\psi}^2}{4B} -B\psi^{\prime 2}
\end{align}
and

\begin{align}
\label{eq:Hr2c}
H_r=0 =-P_{\Lambda}^\prime +P_R + P_{z1}z_{1}^\prime +P_{z2}z_{2}^\prime +P_{\psi}\psi^\prime
\end{align}
for $P_R$ and $\PL$. 

At this point we can consider singularity resolution at the origin in both the metric, \eqref{eq:generalds2}, \eqref{eq:consistency12}, \eqref{eq:gaugechi1} and the higher dimensional generalization of the energy momentum tensor, \eqref{eq:Einsteinuunull1}, \eqref{eq:Einsteinvvnull1} and \eqref{eq:Einsteinuvnull1}.  Consider first the metric: we can set $N$ to be a constant at the origin and so \eqref{eq:gaugechi1} tells us that if $\Sc=\sqrt{l^{n-2}M/j}$ is not singular at $R=0$ then neither is the metric.  So, we can resolve both the singularity in the energy momentum tensor and the one in the metric (in the case where \eqref{eq:condition2} is satisfied) by choosing the functions $V(R)$, $\phi(R)$ and $h(R)$ in the action so that the factors containing $j$ ($=\phi'$) are not singular at the origin.  Note that this choice will not violate the Bianchi identities since we derived our equations of motion from a variational principal.


With the gauge choices, \eqref{eq:chi1} and \eqref{eq:xi1}, the equations of motion for the remaining phase space variables can now be written as

\begin{align}
\label{eq:psiEoM2}
{\dot \psi}=& -N\frac{P_{\psi}}{2B} +N_r \psi^\prime,
\end{align}

\begin{align}
\label{eq:PpsiEoM2}
{\dot P_{\psi}}=& (-2NB\psi^\prime
+ N_r P_\psi )^\prime,
\end{align}

\begin{align}
  \label{eq:z1EoM2}
  {\dot z_1} = -N\Sd -\frac{NP_{z2}}{W}
+N_r z_{1}^\prime,
\end{align}

\begin{align}
  \label{eq:Pz1EoM2}
  {\dot P_{z1}} =& (-W(Nz_{2}^\prime +2N^\prime) +N_r P_{z1} )^\prime
   \nonumber \\
& -NWb( {\Sc}^2 -1 ),
\end{align}

\begin{align}
  \label{eq:z2EoM2}
  {\dot z}_{2} = -N\Sd -\frac{NP_{z1}}{W}
+N_r z_{2}^\prime
\end{align}
and
\begin{align}
  \label{eq:Pz2EoM2}
  {\dot P}_{z2} = (-W(Nz_{1}^\prime +2N^\prime) +N_r P_{z2} )^\prime.
\end{align}

These equations of motion, \eqref{eq:psiEoM2}, \eqref{eq:PpsiEoM2}, \eqref{eq:z1EoM2}, \eqref{eq:Pz1EoM2}, \eqref{eq:z2EoM2}, \eqref{eq:Pz2EoM2} along with the consistency conditions, \eqref{eq:consistency12},  \eqref{eq:gaugechi1}, the constraints, \eqref{eq:H2c}, \eqref{eq:Hr2c} and the definitions, \eqref{eq:S12}, \eqref{eq:S22}, \eqref{eq:S32}, \eqref{eq:Ry3} can be used to evolve appropriate initial conditions forward in time to show the formation of a black hole with a radiating term.

\section{Conclusions}
\label{sec:Conclusions}
In this work we defined an action which mimics spherically symmetric, dimensionally reduced gravity with a radiating term.  Although the radiating term was borrowed from the conformal anomaly and is only valid in two dimensional physics we incorporated terms which account for the non-minimal coupling (in 2-D) of the matter field.  We derived Lagrange's equations in null gauge and found (in agreement with \cite{Ayal1997}) that the energy momentum tensor is singular at the origin.  We found, however, that we could remove this singularity (as well as the singularity in the vacuum solution) by appropriate choice of coefficients in the action without violating the Bianchi identities since our equations of motion were derived from a variational principal.  We then performed a detailed Hamiltonian analysis, including a prescription of the boundary conditions and corresponding boundary terms needed to make the variational principle well defined. From this Hamiltonian we imposed suitable gauge fixing conditions and derived the equations of motion in a family of non-null coordinates.  The fact that these are first order in time derivatives makes them well suited to numerical simulations of black hole formation.  We then chose a gauge such that the metric is well defined at horizon formation, which is well suited to the investigation of the dynamics past horizon formation.

At first glance the Lagrangian \eqref{eq:Action2b} that is our starting point contains many arbitrary functions. However, as we saw, the requirement that singularities be removed is quite restrictive. As shown in Section \ref{sec:EMTensor} in order for the effective stress energy tensor to be regular it is necessary that $j=\omega^2=(\phi')^2$ at $R=0$. In addition, $\phi'$ must go to zero at the origin at least as fast as $R^2$, as can be seen from \eqref{eq:Einsteinnull2}. Finally, the remaining freedom in $j$, $\phi$ and $V$ can be fixed so that the vacuum solution approaches the form derived via general arguments in \cite{Poisson1990}. This leaves the freedom in the matter coupling $B(R)$. 
The simplest choice is $B(R)=R^{n-2}$, i.e. the classical form obtained by dimensional reduction which, as shown in \cite{Ziprick2010} yields non-singular collapse. 
 $b(R)$ and $c(R)$ are normally taken to be zero in two dimensional models, but, when the coupling is not conformal these functions are determined by our choice of $B(R)$.  It would be interesting to see what effect these terms have on the structure of the evaporating, non-singular black hole spacetime. In an upcoming paper we will present the results of numerical simulations using these equations.

\section{Acknowledgements}
We thank Nils Deppe for useful discussions.  This work has been funded by the Fondecyt grant 3140123.  The Centro de Estudios Cient{\'i}ficos (CECs) is funded by the Chilean Government through the Centers of Excellence Base Financing Program of Conicyt. This work was also
funded in part by the Natural Sciences and Engineering Research Council 
of Canada.  Support was also provided by WestGrid (www.westgrid.ca), 
Compute Canada/Calcul Canada (www.computecanada.ca), and the Perimeter 
Institute for Theoretical Physics (funded by Industry Canada and the Province
of Ontario Ministry of Research and Innovation).

\appendix

\section{Loop Quantum Gravity Black Hole}
\label{app:KPBH}

An example of a non-singular, single horizon black hole in four dimensions that can be derived as a solution to \eqref{eq:Action2} is that of \cite{Peltola2009}. See also \cite{Modesto2004} for an earlier treatment.
\bea
ds^2 &=& -\left(\sqrt{1-\frac{k^2}{R^2}}-\frac{l^2M}{2R}\right)dt^2\nonumber\\
 & &+\left(\sqrt{1-\frac{k^2}{R^2}}-\frac{l^2M}{2R}\right)^{-1}\frac{dR^2}{1-\frac{k^2}{R^2}} +R^2 d\Omega^2\nonumber\\
&=&\sqrt{1-\frac{k^2}{R^2}}\left[\left(1-\frac{2GM}{R\sqrt{1-\frac{k^2}{R^2}}}\right)dt^2\right.\nonumber\\
 & &+\left.\left(1-\frac{l^2M}{2R\sqrt{1-\frac{k^2}{R^2}}}\right)^{-1}\left(\frac{dR}{{1-\frac{k^2}{R^2}}}\right)^2\right]\nonumber\\
 & & +R^2 d\Omega^2
\label{eq:PKBH}
\eea
where $k$ is the polymerization (quantum gravity) length scale. This is of the same form as \eqref{eq:ds2vacuum} with the following identifications:
\begin{align}
j(R)&= {2R\sqrt{1-\frac{k^2}{R^2}}}\\
\phi^\prime(R) &= \frac{j(R)}{ 1-\frac{k^2}{R^2}} = \frac{2R}{\sqrt{1-\frac{k^2}{R^2}}}\\
\phi(R)&= \left(R\sqrt{R^2-k^2} + k^2 \ln(R+\sqrt{R^2-k^2}\right)\\
\omega^2 &= \frac{j(R)}{\sqrt{1-\frac{k^2}{R^2}}}={2R}\\
h(R) &= \ln(\omega^2)^\prime\phi^\prime(R)=\frac{\phi^\prime(R)}{R} = \frac{1}{\sqrt{1-k^2/R^2}}\\
V(R)&= \omega^2 \frac{j^\prime(R)}{\phi^\prime} =2
\end{align}
 As shown in \cite{Peltola2009} the above metric can be analytically continued to describe a complete non-singular spacetime containing a single bifurcative horizon.  The areal radius is bounded below by $k$ and re-expands to infinity in the interior of the horizon.  Note that for this black hole the conditions (\ref{eq:condition2}) are not satisfied, so that there are in principle terms in the effective stress tensor that are singular at $R=0$. In the present case this is not an issue because, as previously mentioned, $R=0$ is excluded from the complete, regular spacetime.

\section{Misner-Sharpe Mass Function}
\label{app:mass}

The following analysis describes the mass function used in this work.

\subsection{General Relativity}
In general relativity with no radiation terms the Misner-Sharp mass function is defined by \cite{Taves2013}:
\bea
l^{n-2}M&=&(n-2)R^{n-3} \left(1-(DR)^2\right)\nonumber\\
 &=& (n-2)R^{n-3}\left(1-\left[\Lambda^{-2}R_{,x}^2 -R_{,y}^2\right]\right)
\label{eq:mass function}
\eea
where, from now on we absorb a factor of $l^{n-2}$ into the mass function.  In the case of 4D GR, ie, $\tilde{\phi}=\phi = R^{2}$, $h_z=h = 2 = V$, the mass function is given by
\be
M=2R\left(1-\left[\left(\frac{R_{,x}}{\Lambda}\right)^2 -\left(\frac{P_\Lambda}{4R}\right)^2\right]\right).
\ee
It is then easy to verify that:
\be
-\tilde{H}:=-\frac{R_{,x}}{\Lambda}H - \frac{P_\Lambda}{4\Lambda R}H_r =  M_{,x}
\ee

It therefore makes sense to write the Hamiltonian in terms of the new Hamiltonian constraint:
\be
{\cal H} = \tilde{N} \tilde{H} + \tilde{N}_r H_r= -\tilde{N} M_{,x}+\tilde{N}_r H_r
\ee
with suitably redefined lagrange multipliers:
\begin{align}
\tilde{N}&:=\frac{\Lambda}{R_{,x}}N\\
\tilde{N}_r &:=N_r - \frac{P_\Lambda}{4R R_{,x}}N
\end{align}
Note that with asymptotically flat boundary conditions the only boundary terms that arise in the variation are of the form $\tilde{N} (\delta M)_{,x}$. The boundary term that needs to be added to the Hamiltonian in order to make the variational principle well defined is therefore:
\begin{align}
\label{eq:boundaryterm}
H_B = \tilde{N}_\infty M_\infty
\end{align}
assuming that the mass function vanishes on the inner boundary: $M_{R=0}=0$.  This is discussed for the radiating case in Appendix \ref{app:boundary}. 

\subsection{General $\phi$, $h$ and $V$ with No Radiation}

The mass function for general $\phi$, $h$ and $V$ with no radiating terms is given by
\bea
\label{eq:mass1}
M&=& j\left(1- \frac{(\phi^\prime)^2}{j\omega^2}(DR)^2\right)\nonumber\\
 &=& j\left(1- \frac{(\phi^\prime)^2}{j\omega^2}\left[\left(\frac{R_{,x}}{\Lambda}\right)^2 -\left(\frac{P_\Lambda}{2\phi^\prime}\right)^2\right]\right).
\eea
This is a mass function in the sense that it can be written as a combination of constraints,
\be
-\tilde{H} := -\frac{\phi^\prime R_{,x}}{\omega^2 \Lambda} H - \frac{P_\Lambda}{2\omega^2 \Lambda}H_r = M_{,x}.
\ee
and it reduces to the Misner-Sharp mass function for the appropriate values of $\phi$, $h$ and $V$.  As in the GR case we define
\be
{\cal H} = \tilde{N} \tilde{H} + \tilde{N}_r H_r = -\tilde{N} M_{,x}+\tilde{N}_r H_r
\ee
with suitably redefined lagrangian multipliers:
\begin{align}
\tilde{N}&:=\frac{\omega^2\Lambda}{\phi^\prime R_{,x}}N\\
\tilde{N}_r &:=N_r - \frac{P_\Lambda}{2\phi^\prime R_{,x}}N
\end{align}

\subsection{General $\phi$, $h$ and $V$ with Radiation}

In the most general radiating case that we consider, the derivative of the mass function (\ref{eq:generalM1}) cannot be written 
as a linear combination of the constraints.
In the GR case where $\phi^{\prime \prime} = h = V$ (and therefore ${\phi^\prime}^2/j\omega^2=1$) it is shown in appendix \ref{app:boundary} that the boundary term which must be added to the Hamiltonian is of the form \eqref{eq:boundaryterm} with $M$ given by \eqref{eq:mass function}.  In terms of phase space variables, \eqref{eq:Ry2} can be used to give,
\begin{align}
\label{eq:radmass1}
M
 &= (n-2)R^{n-3} \times \nonumber \\
&\left(1-\left[\left(\frac{R_{,x}}{\Lambda}\right)^2 -\left(\frac{-\phit^\prime \Sa -2W P_R}{\Sb}\right)^2\right]\right)\nonumber\\
\end{align}
The derivative of this mass function also cannot in general be written as a combination of the constraints in the full radiating case.
Nonetheless, the analysis of appendix \ref{app:boundary} applies to the non-GR case as long as the conditions  $\phi^{\prime \prime} = h = V$ are satisfied in the asymptotic region.  In this case, as $R\to\infty$,
\begin{align}
\label{eq:radmass2}
M
 &\to j\left(1-\left[\left(\frac{R_{,x}}{\Lambda}\right)^2 -\left(\frac{-\phit^\prime \Sa -2W P_R}{\Sb}\right)^2\right]\right)\nonumber\\
\end{align}
and this provides the boundary term needed to make the variational principle well defined, as well as the corresponding conserved energy.

\section{Boundary Conditions and Boundary Term}
\label{app:boundary}


Here we derive the boundary conditions at $x \to \infty$ which give a finite Hamiltonian.  For this to happen the Hamiltonian density \eqref{eq:Hamconstraint1} must go to zero faster than $x^{-1}$.  For the non-radiation fields we adopt the boundary conditions used in \cite{Taves2013} and \cite{Kunstatter2013} for slices which approach flat slice co-ordinates as $x \to \infty$,

\begin{align}
\label{eq:Ninf1}
&N\simeq N_\infty(t)+\mathcal{O}(x^{-\epsilon_N}),\\
\label{eq:Nrinf1}
&N_r\simeq N_r^{\infty}(t) x^{-(n-3)/2},\\
\label{eq:Lambdainf1}
&\Lambda \simeq 1,\\
\label{eq:Rinf1}
&R\simeq x+R_1(t) x^{-\epsilon_R},
\end{align}
where $\epsilon_N>0$, $\epsilon_R > 1$ for $n=4$ and $\epsilon_R > n-4$ for $n>4$.  In this derivation we ignore the matter terms which are treated in \cite{Husain2005}.  For the radiation fields we assume the form

\begin{align}
  \label{eq:z1inf1}
  &z_1 \simeq z_{10}^\infty(t) + z_{1}^\infty(t) x^{-\epsilon_{z1}}\\ 
  \label{eq:z2inf1}
  &z_2 \simeq z_{20}^\infty(t) + z_{2}^\infty(t) x^{-\epsilon_{z2}}
\end{align}
We will use the fact that the Hamiltonian should be finite to find the conditions on $z_{10}^\infty$, $z_{20}^\infty$, $\epsilon_{z1}$ and $\epsilon_{z2}$.

In this appendix we assume that the fields $\phi$, $h$, $V$ and $B$ are those of GR case and that $b$ and $c$ go to zero at least as fast as (grow at least as slowly as)

\begin{align}
  \label{eq:b2}
  b=-3(B^\prime/B)^2
\end{align}

\begin{align}
  \label{eq:c2}
  c=-6\ln{B}
\end{align}
although we will discuss the applicability of this analysis to the singularity resolved case of \eqref{eq:phiZip}, \eqref{eq:hZip} and \eqref{eq:VZip}.

We can write down the asymptotic form of the conjugate momenta, up to dominant terms, using \eqref{eq:PLambda1}, \eqref{eq:PR1}, \eqref{eq:Pz11} and \eqref{eq:Pz21}, 

 \begin{align}
   \label{eq:PLinf1}
   \PL \simeq & 2(n-2)N_\infty^{-1}(-R_{1,t} x^{n-3-\epsilon_R} + N_r^\infty x^{(n-3)/2}) \nonumber \\
&-2WN_\infty^{-1}(z_{10,t}^\infty + z_{20,t}^\infty + z_{1,t}^\infty x^{-\epsilon_{z1}} + z_{2,t}^\infty x^{-\epsilon_{z2}}), \\
   \label{eq:PLprimeinf1}
    P_{\Lambda,x} \simeq &  2(n-2)N_\infty^{-1}(-(n-3-\epsilon_R)R_{1,t} x^{n-4-\epsilon_R} \nonumber \\
&+ N_r^\infty ((n-3)/2)x^{(n-5)/2}) \nonumber \\
&+2WN_\infty^{-1}( \epsilon_{z1}z_{1,t}^\infty x^{-\epsilon_{z1}-1} + \epsilon_{z2}z_{2,t}^\infty x^{-\epsilon_{z2}-1}), \\
  \label{eq:PRinf1}
  P_R \simeq &  2(n-2)N_\infty^{-1}(-(n-3)R_{1,t} x^{n-4-\epsilon_R} \nonumber \\
&+ N_r^\infty ((n-3)/2)x^{(n-5)/2}) \nonumber \\
&+6W(n-2)^2N_\infty^{-1}(z_{10}^\infty + z_{1}^\infty x^{-\epsilon_{z1}} -\ln{\mu^2}) \nonumber \\ 
& \times (R_{1,t} x^{-\epsilon_R-2} - N_r^\infty x^{-(n+1)/2}), \\
  \label{eq:Pz1inf1}
  P_{z1} \simeq& -WN_{\infty}^{-1}((n-3)N_r^\infty x^{-(n-1)/2} + z_{20,t}^\infty + z_{2,t}^\infty x^{-\epsilon_{z1}}),  \\
  \label{eq:Pz2inf1}
  P_{z2} \simeq& -WN_{\infty}^{-1}((n-3)N_r^\infty x^{-(n-1)/2} + z_{10,t}^\infty + z_{1,t}^\infty x^{-\epsilon_{z2}}).
 \end{align}

At this point it is advantageous to consider the $NP_{z1}P_{z2}/W\Lambda$ term in \eqref{eq:Hamconstraint1}.  The relevant, radiation terms in this term go as

\begin{align}
\label{eq:PzPzterm}
& z_{10,t}^\infty z_{20,t}^\infty + z_{1,t}^\infty z_{20,t}^\infty x^{-\epsilon_{z1}} \nonumber \\
& + z_{2,t}^\infty z_{10,t}^\infty x^{-\epsilon_{z2}} + z_{1,t}^\infty z_{2,t}^\infty x^{-(\epsilon_{z1} + \epsilon_{z2})},
\end{align}
which are not cancelled by any of the other terms in \eqref{eq:Hamconstraint1}.   The form of \eqref{eq:PzPzterm} means that $\epsilon_{z1} + \epsilon_{z2} > 1$ and $ z_{10,t}^\infty z_{20,t}^\infty = 0$.  In the case where $b=c=0 \to z_1=z_2:=z/2, P_{z1}=P_{z2}:=P_z$ we can show that $z \sim x^{\epsilon_z}$ where $\epsilon_z > 1/2$.  In the spirit of this limit we take

\begin{align}
 \label{eq:zlimits1}
 \epsilon_{z1} > 1/2, \quad  \epsilon_{z2} > 1/2, \quad z_{10}^\infty =0, \quad z_{20}^\infty =0.
\end{align}
Using these conditions and keeping only the biggest terms we then find

 \begin{align}
   \label{eq:PLinf2}
   \PL \simeq & 2(n-2)N_\infty^{-1}(-R_{1,t} x^{n-3-\epsilon_R} + N_r^\infty x^{(n-3)/2}) \\
   \label{eq:PLprimeinf2}
    P_{\Lambda,x} \simeq &  2(n-2)N_\infty^{-1}(-(n-3-\epsilon_R)R_{1,t} x^{n-4-\epsilon_R} \nonumber \\
&+ N_r^\infty ((n-3)/2)x^{(n-5)/2}) \\
  \label{eq:PRinf2}
  P_R \simeq &  2(n-2)N_\infty^{-1}(-(n-3)R_{1,t} x^{n-4-\epsilon_R} \nonumber \\
&+ N_r^\infty ((n-3)/2)x^{(n-5)/2}) \\
  \label{eq:Pz1inf2}
  P_{z1} \simeq& -WN_{\infty}^{-1} z_{2,t}^\infty x^{-\epsilon_{z1}}  \\
  \label{eq:Pz2inf2}
  P_{z2} \simeq& -WN_{\infty}^{-1} z_{1,t}^\infty x^{-\epsilon_{z2}}.
 \end{align}
Plugging these and \eqref{eq:zlimits1} into \eqref{eq:H2b} and \eqref{eq:Hr2b} and dropping all non-dominant terms we can see that the Hamiltonian density reduces to the non-radiating case and the rest of the proof that the Hamiltonian is finite can be found in Appendix A.2 of \cite{Taves2013}.

It is important to note that there are some cancellations in the rest of the derivation which require $2\phi^{\prime \prime} - h_z - V \to 0$ as $x \to \infty$.  This occurs in the GR case as well as for \eqref{eq:phiprimeZip}, \eqref{eq:hZip} and \eqref{eq:VZip} assuming that $b$ and $c$ go to zero fast enough.  These must be satisfied in order to use the boundary conditions given by \eqref{eq:Ninf1}, \eqref{eq:Nrinf1}, \eqref{eq:Lambdainf1} and \eqref{eq:Rinf1}.

We now take the variation of the action to find which boundary terms do not approach zero as $x \to \infty$.  The variation of the action is given by

\begin{align}
\delta I=&\int dt\int dx \biggl( \partial_t (\delta {\Lambda}P_{\Lambda})-\delta \Lambda{\dot P_{\Lambda}} + {\dot \Lambda}\delta P_{\Lambda}  \nonumber \\
&+ \partial_t (\delta {R}P_{R}) -\delta R{\dot P_{R}}+{\dot R}\delta P_{R} \nonumber \\
&+ \partial_t (\delta {z_1}P_{z1}) -\delta z_1{\dot P_{z1}}+{\dot z}_1\delta P_{z1} \nonumber \\
&+ \partial_t (\delta {z_2}P_{z2}) -\delta z_2{\dot P_{z2}}+{\dot z}_2\delta P_{z2} \nonumber \\
& -\delta N H-N \delta H-\delta N_r H_{r}-N_r \delta H_{r}\biggl),\label{variation1gr}
\end{align}
Assuming that all variations vanish at the time end points the only contributions to the boundary term come from the last line of \eqref{variation1gr}.  Using the boundary conditions \eqref{eq:Ninf1}, \eqref{eq:Nrinf1}, \eqref{eq:Lambdainf1}, \eqref{eq:Rinf1}, \eqref{eq:z1inf1}, \eqref{eq:z2inf1} and \eqref{eq:zlimits1} we calculate the boundary term.  This calculation is tedious but straight forward and can be found in appendix A.2 of \cite{Taves2013} for the non-radiating case.  The only boundary term which does not approach zero as $x \to \infty$ is

\begin{align}
  \label{eq:boundaryterm1}
  N_r \Lambda \delta P_\Lambda \simeq & N_r^\infty x^{-(n-3)/2}[2(n-2)\delta(N_r^\infty/N_\infty)x^{(n-3)/2}] \nonumber \\
 =& (n-2)N_\infty \delta(N_r^{\infty 2} /N_\infty^2).
\end{align}
The Misner-Sharp mass in this case (see (1.35) of \cite{Taves2013}), with $l$ set to 1, is given by

\begin{align}
  \label{eq:MS1}
  M=(n-2)R^{n-3}(N_r/N)^2
\end{align}
which goes to $(n-2)(N_r^\infty / N_\infty)^2$ as $x \to \infty$ and so the the variation of the action is

\begin{align}
  \label{eq:actionvariation2}
  \delta I = \int dt dx (\mbox{dynamical terms}) + \int dt \biggl[N \delta M\biggl]_{x=-\infty}^{x=+\infty}.
\end{align}
This requires the addition of $NM|_{R=\infty}$ to the Hamiltonian.

\bibliography{library}

\begin{thebibliography}{10}

\bibitem{Almheiri2013a}
Ahmed Almheiri, Donald Marolf, Joseph Polchinski, Douglas Stanford, and James
  Sully.
\newblock {An apologia for firewalls}.
\newblock {\em Journal of High Energy Physics}, 2013(9):18, September 2013.

\bibitem{Almheiri2013}
Ahmed Almheiri, Donald Marolf, Joseph Polchinski, and James Sully.
\newblock {Black Holes: Complementarity or Firewalls?}
\newblock {\em Journal of High Energy Physics}, 2013(2):1--20, 2013.

\bibitem{Vaz2014}
Cenalo Vaz.
\newblock {Black Holes as Gravitational Atoms}.
\newblock {\em International Journal of Modern Physics D}, 23(12):1441002,
  2014.

\bibitem{Vaz2014a}
Cenalo Vaz.
\newblock {There's Nothing "Black" about a Black Hole}.
\newblock {\em arXiv preprint arXiv:1407.3823}, 2014.

\bibitem{Frolov1981}
Valeri~P. Frolov and G.A. Vilkovisky.
\newblock {Spherically Symmetric Collapse in Quantum Gravity}.
\newblock {\em Physics Letters B}, 106(4):307--313, 1981.

\bibitem{Frolov1979}
Valeri~P. Frolov and G.~A. Vilkovisky.
\newblock {Quantum Gravity Removes Classical Singularities and Shortens the
  Life of Black Holes}.
\newblock {\em Triest preprint IC-79-69}, November 1979.

\bibitem{Ashtekar2005}
Abhay Ashtekar and Martin Bojowald.
\newblock {Black Hole Evaporation: A Paradigm}.
\newblock {\em Classical and Quantum Gravity}, 22(16):3349, 2005.

\bibitem{Varadarajan2008}
Madhavan Varadarajan.
\newblock {Quantum Gravity and the Information Loss Problem}.
\newblock {\em Journal of Physics: Conference Series}, 140:012007, November
  2008.

\bibitem{Hayward2006}
Sean~A. Hayward.
\newblock {Formation and Evaporation of Non-singular Black Holes}.
\newblock {\em Physical Review Letters}, 96:031103, 2006.

\bibitem{Ziprick2009}
Jonathan Ziprick.
\newblock {\em {Singularity Resolution and Dynamical Black Holes}}.
\newblock Msc, Manitoba, 2009.

\bibitem{Ziprick2010}
Jonathan Ziprick and Gabor Kunstatter.
\newblock {Quantum Corrected Spherical Collapse: A Phenomenological Framework}.
\newblock {\em Physical Review D}, 82(4):044031, 2010.

\bibitem{Hossenfelder2010}
Sabine Hossenfelder, Leonardo Modesto, and Isabeau Pr\'{e}mont-Schwarz.
\newblock {Model for Nonsingular Black Hole Collapse and Evaporation}.
\newblock {\em Physical Review D}, 81(4):044036, 2010.

\bibitem{Zhang2014}
Yiyang Zhang, Yiwei Zhu, Leonardo Modesto, and Cosimo Bambi.
\newblock {Can Static Regular Black Holes Form from Gravitational Collapse?}
\newblock {\em arXiv preprint arXiv:1404.4770}, 2014.

\bibitem{Roman1983}
Thomas Roman and Peter Bergmann.
\newblock {Stellar Collapse without Singularities?}
\newblock {\em Physical Review D}, 28(6):1265--1277, 1983.

\bibitem{Polyakov1981}
AM~Polyakov.
\newblock {Quantum geometry of bosonic strings}.
\newblock {\em Physics Letters B}, 103(3):207--210, 1981.

\bibitem{Callan1992}
Curtis~G. Callan, Steven~B. Giddings, Jeffrey~A. Harvey, and Andrew Strominger.
\newblock {Evanescent Black Holes}.
\newblock {\em Physical Review D}, 45(4):R1005--R1010, 1992.

\bibitem{Russo1992a}
Jorge Russo, Leonard Susskind, and L\'{a}rus Thorlacius.
\newblock {End Point of Hawking Radiation}.
\newblock {\em Physical Review D}, 46(8):3444--3449, 1992.

\bibitem{Ashtekar2011}
Abhay Ashtekar, Frans Pretorius, and Fethi Ramazanoğlu.
\newblock {Evaporation of Two-Dimensional Black Holes}.
\newblock {\em Physical Review D}, 83(4):1--18, 2011.

\bibitem{Ashtekar2011a}
Abhay Ashtekar, Frans Pretorius, and Fethi Ramazanoğlu.
\newblock {Surprises in the Evaporation of 2D Black Holes}.
\newblock {\em Physical Review Letters}, 106(16):1--4, 2011.

\bibitem{Ayal1997}
Shai Ayal and Tsvi Piran.
\newblock {Spherical Collapse of a Massless Scalar Field with Semiclassical
  Corrections}.
\newblock {\em Physical Review D}, 56(8):4768--4774, 1997.

\bibitem{Parentani1994}
Renaud Parentani and Tsvi Piran.
\newblock {Internal Geometry of an Evaporating Black Hole}.
\newblock {\em Physical Review Letters}, 73(21):2805--2808, 1994.

\bibitem{Frolov2014}
Valeri~P. Frolov.
\newblock {Information Loss Problem and a ‘Black Hole’ Model with a Closed
  Apparent Horizon}.
\newblock {\em Journal of High Energy Physics}, 2014(5):49, 2014.

\bibitem{Bardeen2014}
JM~Bardeen.
\newblock {Black Hole Evaporation without an Event Horizon}.
\newblock {\em arXiv preprint arXiv:1406.4098}, 2014.

\bibitem{Torres2014}
R.~Torres and F.~Fayos.
\newblock {Singularity Free Gravitational Collapse in an Effective Dynamical
  Quantum Spacetime}.
\newblock {\em Physics Letters B}, 733:169--175, 2014.

\bibitem{Mersini-Houghton2014}
Laura Mersini-Houghton and Harald~P. Pfeiffer.
\newblock {Back-Reaction of the Hawking Radiation Flux on a Gravitationally
  Collapsing Star II: Fireworks Instead of Firewalls}.
\newblock {\em arXiv preprint arXiv:1409.1837}, page~9, September 2014.

\bibitem{Maeda2008}
Hideki Maeda and Masato Nozawa.
\newblock {Generalized Misner-Sharp Quasi-local Mass in Einstein-Gauss-Bonnet
  Gravity}.
\newblock {\em Physical Review D}, 77(6):064031, 2008.

\bibitem{Maeda2011}
Hideki Maeda, Steven Willison, and Sourya Ray.
\newblock {Lovelock Black Holes with Maximally Symmetric Horizons}.
\newblock {\em Classical and Quantum Gravity}, 28(16):165005, 2011.

\bibitem{Kunstatter2012}
Gabor Kunstatter, Tim Taves, and Hideki Maeda.
\newblock {Geometrodynamics of Spherically Symmetric Lovelock Gravity}.
\newblock {\em Classical and Quantum Gravity}, 29(9):092001, 2012.

\bibitem{Kunstatter2013}
Gabor Kunstatter, Hideki Maeda, and Tim Taves.
\newblock {Hamiltonian Dynamics of Lovelock Black Holes with Spherical
  Symmetry}.
\newblock {\em Classical and Quantum Gravity}, 30(6):065002, 2013.

\bibitem{Taves2013}
Tim Taves.
\newblock {\em {Black Hole Formation in Lovelock Gravity}}.
\newblock Phd, University of Manitoba, 2013.

\bibitem{Grumiller2002}
D.~Grumiller, W.~Kummer, and D.~V. Vassilevich.
\newblock {Dilaton Gravity in Two Dimensions}.
\newblock {\em Physics Reports}, 369(4):327--430, 2002.

\bibitem{Gegenberg2010}
Jack Gegenberg and Gabor Kunstatter.
\newblock {2-D Midisuperspace Models for Quantum Black Holes}.
\newblock In Daniel Grumiller, Anton Rebhan, and Dimitri Vassilevich, editors,
  {\em Fundamental Interactions: A Memorial Volume for Wolfgang Kummer}, pages
  231--247. World Scientific, 2010.

\bibitem{Louis-Martinez1994}
Domingo Louis-Martinez and Gabor Kunstatter.
\newblock {Birkhoff's Theorem in Two-Dimensional Dilaton Gravity}.
\newblock {\em Physical Review D}, 49(10):5227--5230, 1994.

\bibitem{Poisson1990}
Eric Poisson and Werner Israel.
\newblock {Internal Structure of Black Holes}.
\newblock {\em Physical Review D}, 41(6):1796----1809, 1990.

\bibitem{Modesto2004}
Leonardo Modesto.
\newblock {Disappearance of the Black Hole Singularity in Loop Quantum
  Gravity}.
\newblock {\em Physical Review D}, 70(12):124009, 2004.

\bibitem{Peltola2009}
Ari Peltola and Gabor Kunstatter.
\newblock {Effective Polymer Dynamics of D-Dimensional Black Hole Interiors}.
\newblock {\em Physical Review D}, 80:044031, 2009.

\bibitem{Nojiri1998}
Shin'ichi Nojiri and Sergei~D. Odintsov.
\newblock {Trace Anomaly Induced Effective Action for 2D and 4D Dilaton Coupled
  Scalars}.
\newblock {\em Physical Review D}, 57(4):2363, 1998.

\bibitem{Kummer1998}
W.~Kummer, H.~Liebl, and D.~V. Vassilevich.
\newblock {Comment on “Trace Anomaly of Dilaton-Coupled Scalars in Two
  Dimensions”}.
\newblock {\em Physical Review D}, 58(10):108501, 1998.

\bibitem{Medved1999}
Joseph Medved and Gabor Kunstatter.
\newblock {Quantum Corrections to the Thermodynamics of Charged 2-D Black
  Holes}.
\newblock {\em Physical Review D}, 60(10):104029, 1999.

\bibitem{Medved2001}
Joseph Medved.
\newblock {\em {Thermodynamics of Charged Back Holes in Two-Dimensional Dilaton
  Gravity}}.
\newblock PhD thesis, University of Manitoba, 2001.

\bibitem{Garfinkle1995}
David Garfinkle.
\newblock {Choptuik Scaling in Null Coordinates}.
\newblock {\em Physical Review D}, 51(10):5558, 1995.

\bibitem{Hayward1995}
J.~D. Hayward.
\newblock {Entropy in the Russo-Susskind-Thorlacius Model.}
\newblock {\em Physical review D: Particles and fields}, 52(4):2239--2244,
  1995.

\bibitem{Gegenberg2012}
Jack Gegenberg, Gabor Kunstatter, and Tim Taves.
\newblock {Quantum Mechanics of the Interior of Radiating 2D Black Holes}.
\newblock {\em Physical Review D}, 85(2):024025, 2012.

\bibitem{Arnowitt2004}
R.~Arnowitt, S.~Deser, and Charles~W. Misner.
\newblock {The Dynamics of General Relativity}.
\newblock {\em Gravitation: An Introduction to Current Research}, pages
  227--264, 1962.

\bibitem{Husain2005}
Viqar Husain and Oliver Winkler.
\newblock {Flat Slice Hamiltonian Formalism for Dynamical Black Holes}.
\newblock {\em Physical Review D}, 71(10):104001, 2005.

\end{thebibliography}
\bibliographystyle{unsrt} 

\end{document}